\documentclass[11pt]{article}
\usepackage{graphicx}
\begin{document}

\title{Marine Le Pen can breach her glass ceiling: \\  The drastic effect of differentiated abstention}

\author{Serge Galam\thanks{serge.galam@sciencespo.fr} \\ CEVIPOF - Centre for Political Research, \\ Sciences Po and CNRS,\\
 98 rue de l'Universit\'e Paris, 75007, France}
 
\date{February 22, 2017}
\maketitle

\begin{abstract}

Ranges of differentiated abstention are shown to reverse an ``exact" poll estimate on voting day allowing the minority candidate to win the election. In a two-candidate competition A and B with voting intentions at $I_a$, $I_b=1-I_a$ and respective turnout at $x$ and $y$,  there exists a critical value $I_{ac}$ for which $I_{ac}<I_a<\frac{1}{2}$ yields an actual election outcome $v_a>\frac{1}{2}$. The reversal may occur without any change of individual choices. Accordingly, for a set of turnouts $x$ and $y$ the minimum voting intention $I_{ac}$ required for A to win the final vote can be calculated. The various ranges of $x$ and $y$ for which $I_{Ac}$ is smaller than the expected 50\% barrier are determined. The calculations are applied to the coming 2017 French presidential election for a second round scenario involving the National Front candidate Marine Le Pen against either the Right candidate Fran\c{c}ois Fillon or the Center candidate Emmanuel Macron. Several realistic conditions are found to make Marine Le Pen win the election despite voting intentions about only 40-45\%.

\end{abstract}


%%%%%%%%%%%%%%%%%%%%%%%%%%%%%%%%%

French Presidential elections are characterized by a two-round voting system. The first round of the 2017 election will be held on April 23 and the second round on May 3. This upcoming election is of a very particular nature combining an unpredictable winner and a predictable loser. While Marie Le Pen should take first place in the first round and thus qualify for the second round, she is expected to be defeated in the second round, whoever her challenger might be. The French presidential electoral system thus exhibits paradoxical features pointing to a blatant non-democratic drawback which ensures the candidate who comes second in the first round will come first in the second round when a National Front candidate is present. Actual race thus resumes to win the second place at the first round.

The expected Le Pen defeat in the second and final round is rooted in the existence of a so called ``Republican Front", which has been activated regularly with total success, besides very rare exceptions, each time a National Front candidate has run in the second round of a local or national election. The Republican Front results from the interplay of two effects. The first effect stems from the refusal of all political parties, Right and Left, to join forces with FN candidates for second round of local elections. The second effect emerges from the adamant refusal of millions of voters to allow a National Front candidate to be elected. In order to prevent this from happening they vote massively to support the challenger candidate regardless of their political affiliation. This creates what has been defined as a ``glass ceiling", which prevents any National Front candidate who runs at the second round from exceeding the required threshold of 50\% of the ballot needed to win the election. Although this Republican Front has eroded substantially over past elections, it continues to maintain the glass ceiling positioned below 50\%. Therefore Marine Le Pen cannot win the second and final round however high her score is since in all cases this score will be below 50\%. In contrast, the National Front did manage to get numerous candidates elected to the European parliament since these elections are proportional \cite{wiki}.

Nevertheless, it is important to stress that the current campaign has been rife with unexpected outcomes embedded with a series of ongoing judicial incidents. Primaries were held successively by the Right-Center and latter by the Left. The outcome of these primaries for both the Left and the Right-Center was the defeat of the favorite candidate on both sides. On the Right (Les R\'epublicains) Alain Jupp\'e was defeated by Fran\c{c}ois Fillon and on the Left, (the Socialist Party) Manuel Valls by Beno\^it Hamon.  As a result, the possibility that the erosion of the Republican Front will accelerate suddenly cannot be dismissed. Although very unlikely to happen, the existence of such a possibility makes the likelihood that Marine Le Pen could be elected shift from impossible to improbable \cite{semidoc}.

While above conclusion results from the usual analysis of National Front dynamics, in this paper I suggest a novel phenomenon which may well shake drastically this existing situation. Indeed, certain ranges of differentiated abstention in the second round of the elections are shown to reverse the expected failure of Marine Le Pen into success without any change in individual choices. On this basis, the likelihood of  Marine Le Pen being elected President of France in 2017 shifts from improbable to quite possible.

To substantiate my claim I develop a simple generic study which departs from classical studies of abstention within political sciences \cite{1,2,3,4,5,6,7}. More generally, I consider a two-candidate competition A and B with ``exact" voting intentions $I_a$ and $I_b=1-I_a$. When $I_a<I_b$, i.e., $I_a<\frac{1}{2}$, some range of differentiated abstention is found to reverse the expected voting order with actual vote outcome $v_a>\frac{1}{2}$ for A and $v_b=1-v_a<\frac{1}{2}$. Indeed, given respective turnout $x$ and $y$,  there exists a critical value $I_{ac}$ for which $I_{ac}<I_a<\frac{1}{2}$ yields $v_a>\frac{1}{2}$. The various ranges of $x$ and $y$ for which $I_{ac}$ is smaller than the 50\% barrier are determined. 

At every election, once voting is completed, three quantities $V_a$, $V_b$, $T$ are obtained, respectively ballots for A, ballots for B, and actual turnout. We thus have $V_a + V_b + (1-T)=1$ since $(1-T)$ measures actual abstention. Blank and null ballots could be accounted for but without changing the results. To simplify the presentation every voter is assumed to have made a choice on voting day, which does not imply to cast a ballot. Usually, these data are rescaled so that the winner is elected by more than $50\%$ of the ballots cast with
\begin{equation} 
V_{a,b} \rightarrow v_{a,b} \equiv \frac{V_{a,b} }{V_a+V_b} ,
\label{v1} 
\end{equation}
with $v_a+v_b=1$. We thus have $v_a$, $v_b$, $T$ instead of $V_a$, $V_b$, $T$ making the winner elected with $v>\frac{1}{2}$.

Respective turnout $x$ and $y$ for A and B correspond to differentiated abstention $(1-x)$ and $(1-y)$ which yield actual turnout
\begin{equation} 
T=x I_a + y I_b = (x-y)I_a+y .
\label{t} 
\end{equation}
It should be stressed that only $T$ is known while $x$ and $y$ are not. In addition, from ``exact" voting intentions $I_a$ and $I_b$ we obtain $V_a=x I_a $ and $V_b= y I_b $. Eq. (\ref{v1}) thus writes
\begin{equation} 
v_{a,b} = \frac{ x I_{a,b}  }{x  I_a +  y I_b }  . 
\label{v2} 
\end{equation}
From Eq. (\ref{v2})  A wins the election when $v_a>v_b \Leftrightarrow x I_a>y I_b=y(1-I_a)$, which yields a critical value for voting intentions for A 
\begin{equation} 
I_{ac} =\frac{y}{x +  y }  . 
\label{i} 
\end{equation}
When $I_a>I_{ac}$ A wins the election with $v_a>v_b$ even if $I_{a} <\frac{1}{2}$. Therefore knowing $x$, $y$ and $I_a$ allows to predict the outcome of the election.  At the critical voting intention $I_a=_{ac}$ the associated critical turnout value writes
\begin{equation} 
T_c= \frac{2x y}{x +  y } .
\label{t} 
\end{equation}

\begin{figure}[t]
\centering
\includegraphics[width=.80\textwidth]{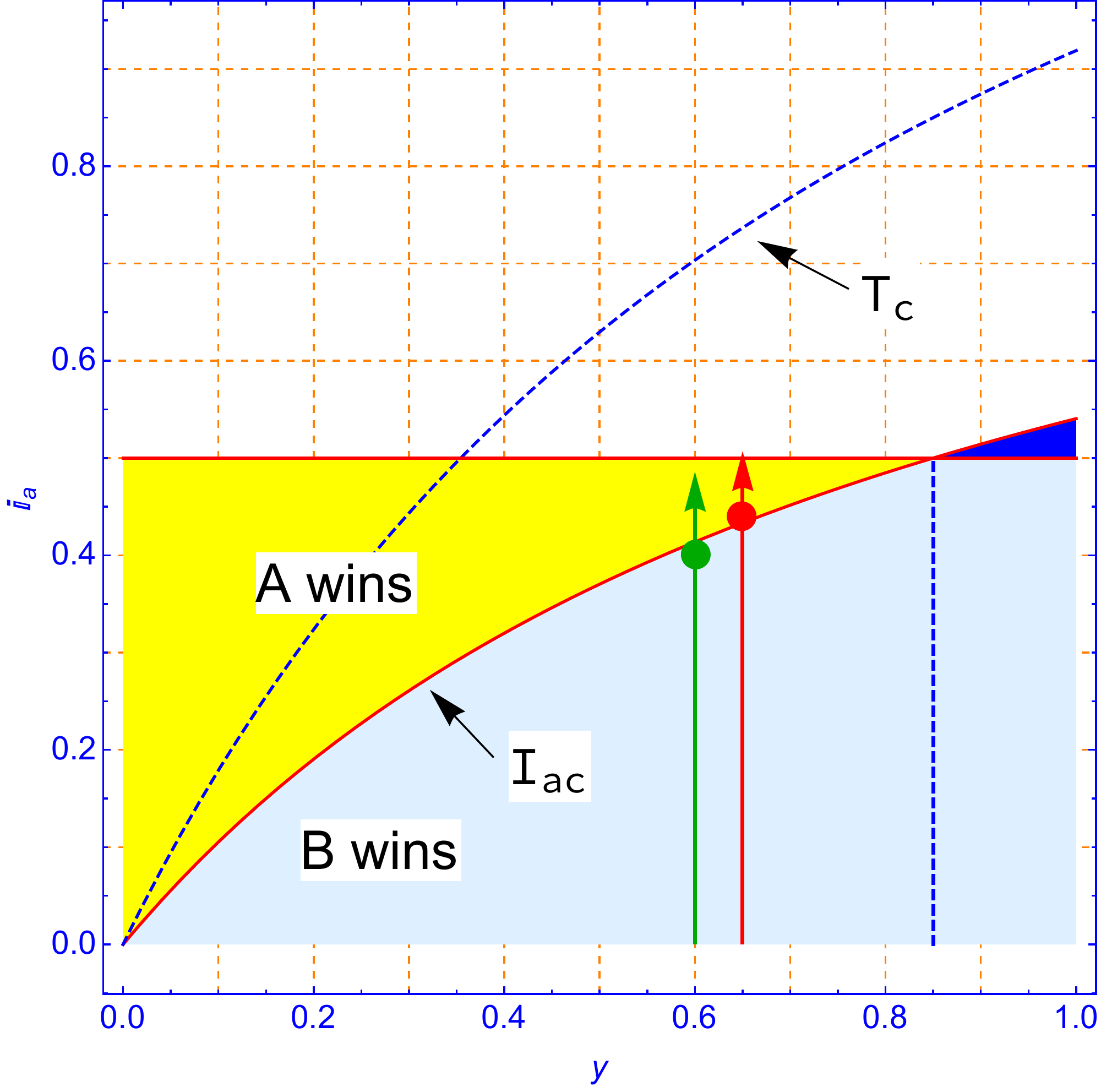}
\caption{The critical line $I_{ac} =\frac{y}{x +  y} $ (lower line, red) is shown as a function of $0\leq y \leq 1$ for $x=0.85$. In the $I_a<I_{ac}$ area (blue lower dark part under the curve) B wins the election. In the $I_a>I_{ac}$  area (upper yellow clear part above the curve) A wins the election. The arrow (red, right side) shows the A vote at $v_a=0.5068$ for $y=0.65$ and $I_a=0.44> I_{ac}=0.4333$ allowing A to win with an ``exact" minority of voting intentions. The dot (red) locates A voting intention at $y=0.65$ for $I_a=0.44$. The arrow (green, left side) shows the the A vote at $v_a=0.4857$ for $y=0.60$ and $I_a=0.40< I_{ac}=0.4138$ not allowing A to win with an ``exact" minority of voting intentions. The dot (green) locates A voting intention at $I_a=0.40$ for $y=0.60$. The critical line $T_{c} =\frac{2x y}{x +  y } $ is also shown (upper line, blue dashed) as a function of $0\leq y \leq 1$ for $x=0.85$.}
\label{i1}
\end{figure}

In Figure (\ref{i1}) the critical line $I_{ac} =\frac{y}{x +  y} $ (red) is shown as a function of $0\leq y \leq 1$. In the $I_a<I_{ac}$ area (blue lower dark part under the curve) B wins the election. In the $I_a>I_{ac}$  area (upper yellow clear part above the curve) A wins the election. The arrow (red, right side) shows the A vote at $v_a=0.5068$ for $y=0.65$ and $I_a=0.44> I_{ac}=0.4333$ allowing A to win with an ``exact" minority of voting intentions. The dot (red) locates A voting intention at $y=0.65$ for $I_a=0.44$. The arrow (green, left side) shows the A vote at $v_a=0.4857$ for $y=0.60$ and $I_a=0.40< I_{ac}=0.4138$ not allowing A to win with an ``exact" minority of voting intentions. The dot (green) locates A voting intention at $I_a=0.40$ for $y=0.60$. 

\begin{figure}[t]
\centering
\includegraphics[width=1\textwidth]{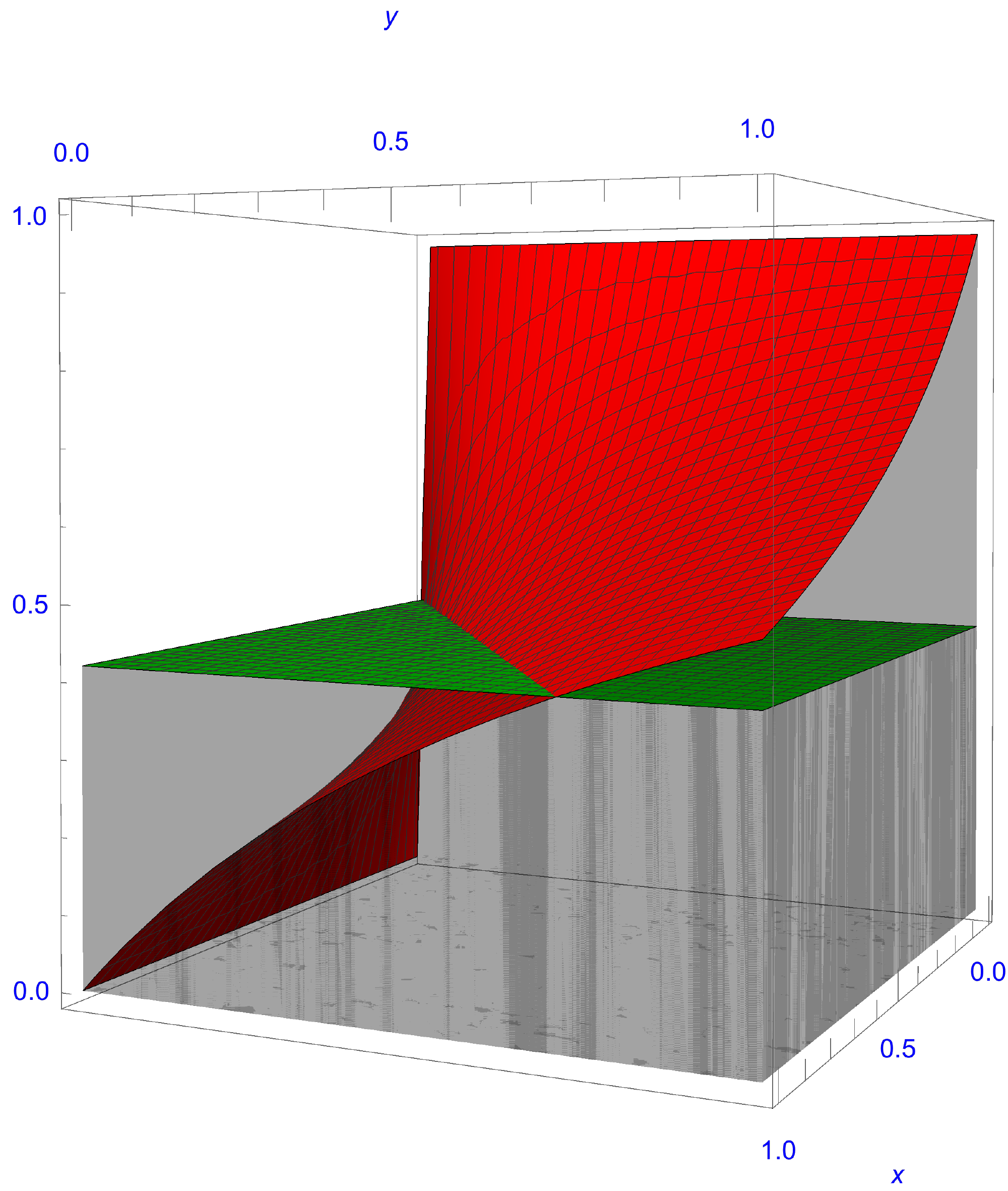}
\caption{The critical surface $I_{ac} =\frac{y}{x +  y}$ is shown as a function of $0\leq x \leq 1$ and $0\leq y \leq 1$. As an example, the intersection with the $I_a=0.42$ (green) plane is exhibited. Part of the plane below the $I_{ac}$ surface (at the back of the graph) leads to a B victory while the part above (at the front) leads to a A victory.}
\label{i2}
\end{figure}

Figure (\ref{i2}) shows the critical surface $I_{ac} =\frac{y}{x +  y} $ a function of $0\leq x \leq 1$ and $0\leq y \leq 1$. As an example, the intersection with the $I_a=0.42$ (green) plane is exhibited. Part of the plane below the $I_{ac}$ surface (at the back of the graph) leads to a B victory while the part above (at the front) leads to a A victory.

Another critical value can be determined for B turnout $y$ with 
\begin{equation} 
y_{c} =\frac{x I_a}{1-I_a }  ,
\label{y} 
\end{equation}
leading to A being elected in the range $y<y_c$ for a set $x$ and $I_a$. In such a case $v_a>\frac{1}{2}$ even if $I_{a} <\frac{1}{2}$.

\begin{figure}[t]
\centering
\includegraphics[width=.80\textwidth]{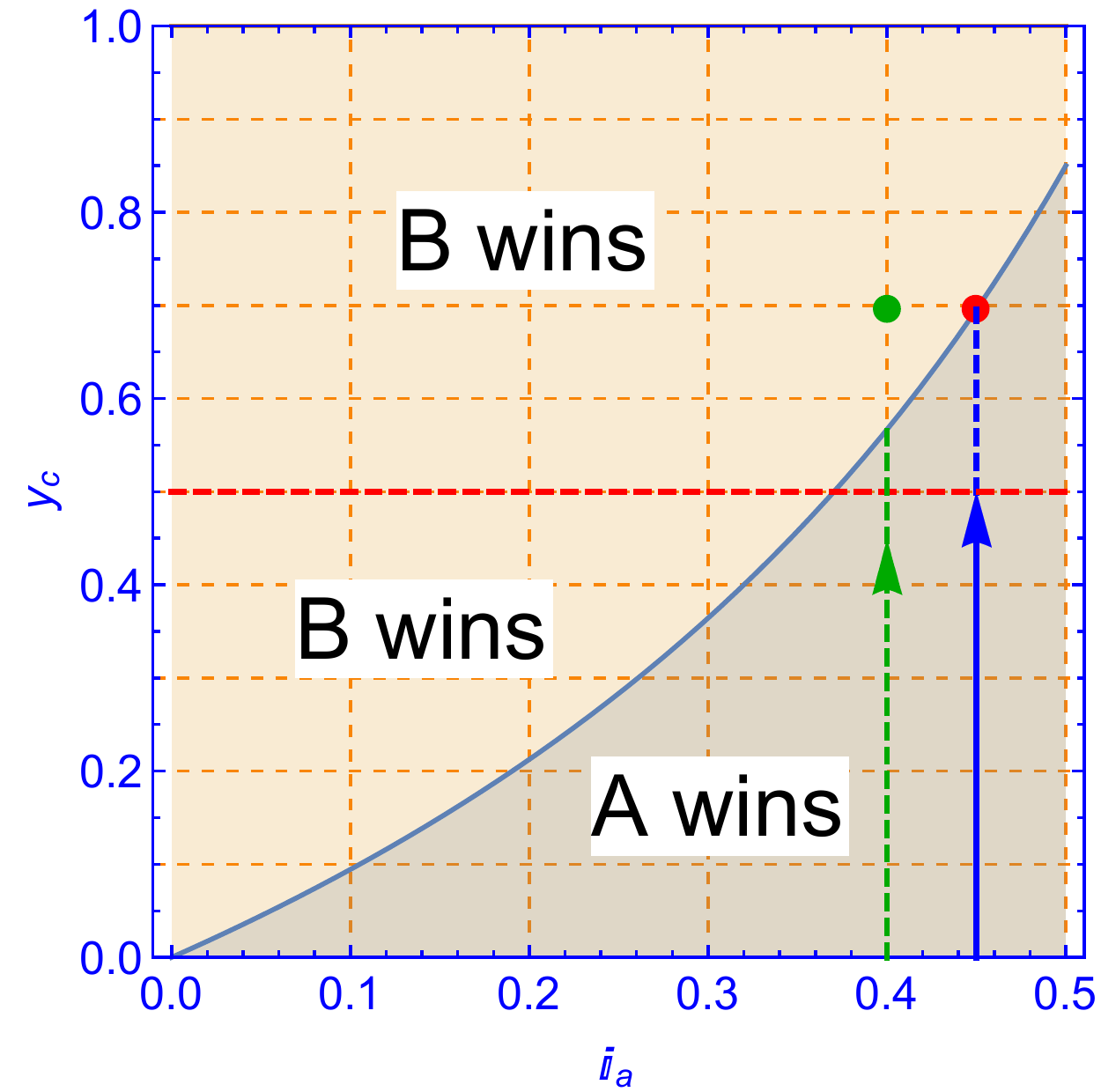}
\caption{The critical line $y_c=\frac{x I_a}{1-I_a }$ is shown as a function of $0\leq I_a \leq \frac{1}{2}$ for $x=0.90$. In the $y<y_c$ area (lower dark part under the curve) A wins the election. In the $y>y_c$ area (upper clear part above the curve) B wins the election. The arrow (blue) shows the A vote at 0.5007 for $I_a=0.42$ and B turnout $y=0.65$ allowing A to win with an ``exact" minority of voting intentions. The dot (red) locates B turnout $y=0.65<y_c=0.6517$ and voting intentions $I_a=0.42$. The arrow (green, left side) shows the A vote at $v_a=0.4491$ for $I_a=0.40$ and B turnout $y=0.695$ not allowing A to win with an ``exact" minority of voting intentions. The dot (green) locates B turnout at $y=0.695>y_c=0.5667$ for $I_a=0.45$.}
\label{y1}
\end{figure}

\begin{figure}[t]
\centering
\includegraphics[width=.80\textwidth]{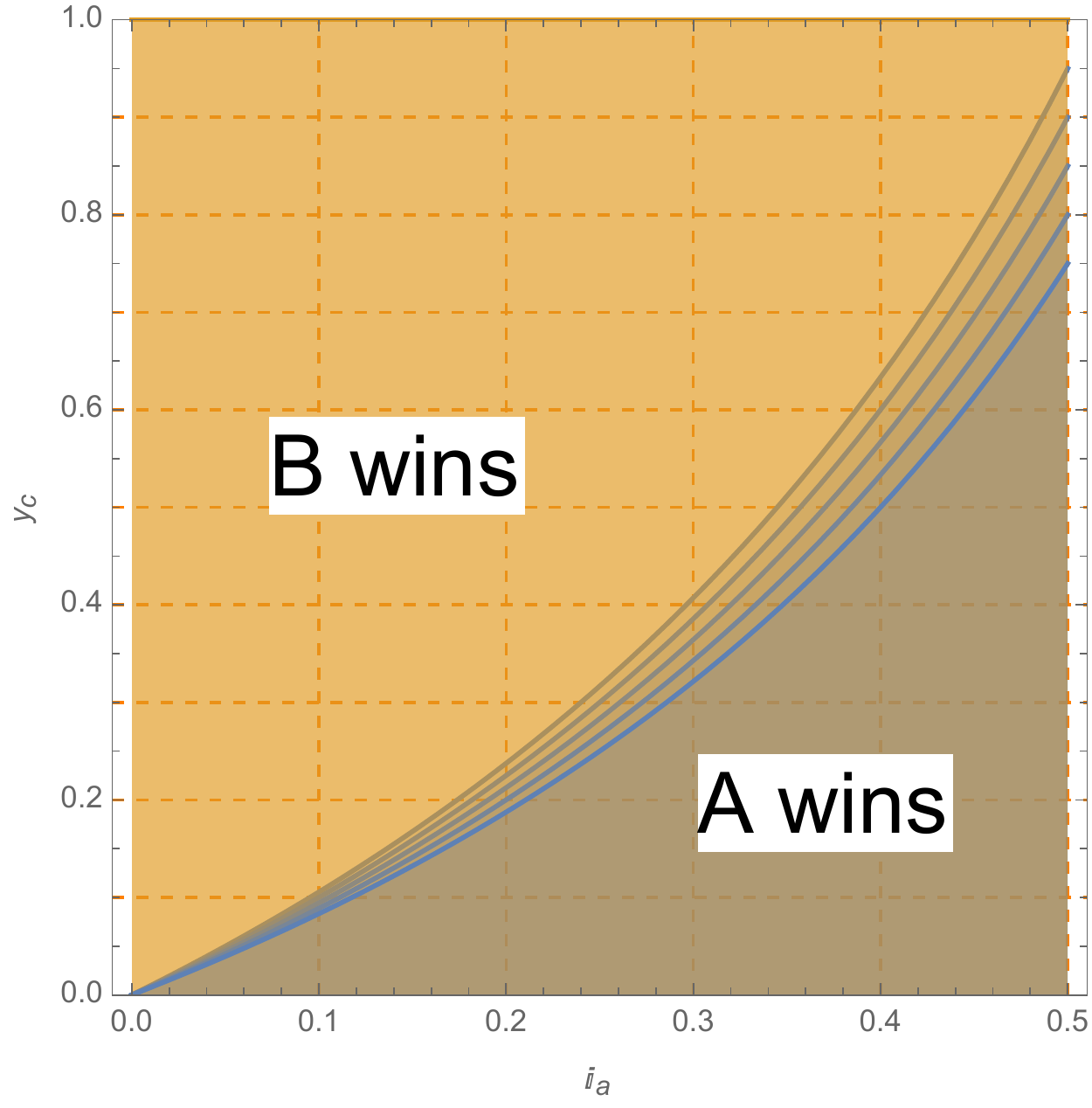}
\caption{The critical line $y_c=\frac{x I_a}{1-I_a }$ is shown as a function of $0\leq I_a \leq \frac{1}{2} $ from top down for $x=0.95,0.90,0.85,0.80,0.75$. In the $y<y_c$ area (lower dark part under the curve) A wins the election. In the $y>y_c$ area (upper clear part above the curve) B wins the election.}
\label{y2}
\end{figure}

\begin{figure}[t]
\centering
\includegraphics[width=1.2\textwidth]{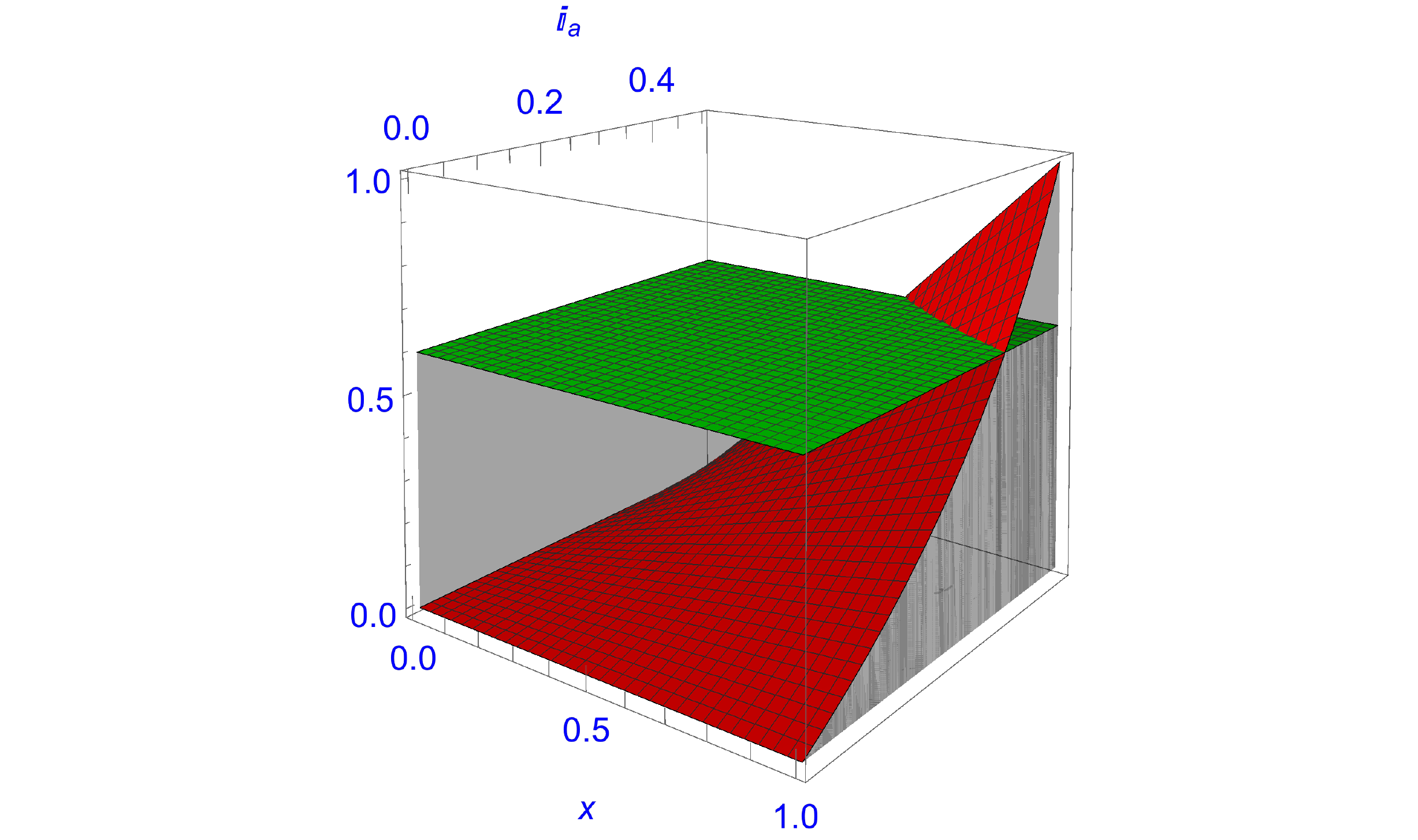}
\caption{The critical surface $y_c=\frac{x I_a}{1-I_a }$ (red) is shown as a function of $0\leq I_a \leq \frac{1}{2}$ and $0\leq x \leq 1$. Intersection with the $y=0.60$ (green) plane is exhibited. Part of the plane below the  $y_c$ surface (at the back of the graph) leads to an A victory while the part above (at the front) leads to a B victory.}
\label{y3}
\end{figure}

However, given that $x$ and $y$ are not known,  Eqs. (\ref{i}) and  (\ref{y}) can only be used  to define ranges of differentiated turnouts which yield the A victory as shown in Figures ( \ref{i1}, \ref{y1}, \ref{y2}, \ref{y3}). It is thus possible to signal when voting intentions are located in turnout ranges for which an unexpected outcome that contradicts poll predictions becomes feasible.

In Figure (\ref{y1}) the critical line $y_c=\frac{x I_a}{1-I_a }$ is shown as a function of $0\leq I_a \leq  \frac{1}{2}$ for $x=0.85$. In the $y<y_c$ area (lower dark part under the curve) A wins the election. In the $y>y_c$ area (upper clear part above the curve) B wins the election. The arrow (blue, right side) shows the A vote at $v_a=0.5002$ for $I_a=0.45$ and B turnout $y=0.695$ allowing A to win with an ``exact" minority of voting intentions. The dot (red) locates B turnout at $y=0.695<y_c=0.7647$ for $I_a=0.45$. The arrow (green, left side) shows the A vote at $v_a=0.4491$ for $I_a=0.40$ and B turnout $y=0.695$ not allowing A to win with an ``exact" minority of voting intentions. The dot (green) locates B turnout at $y=0.695>y_c=0.5667$ for $I_a=0.45$.

Figure (\ref{y2}) shows the variation of the critical curve $y_c=\frac{x I_a}{1-I_a }$ for $x=0.95,0.90,0.85,0.80,0.75$ as a function of $0\leq I_a \leq  \frac{1}{2}$. As an example, the intersection with the $y=0.60$ (green) plane is exhibited. Part of the plane below the $y_c$ surface (at the back of the graph) leads to an A victory while the part above (at the front) leads to a B victory.

Figure (\ref{y3}) is three-dimensional exhibiting the critical surface $y_c=\frac{x I_a}{1-I_a }$ (red)  as a function of $0\leq I_a \leq  \frac{1}{2}$ and $0\leq x \leq 1$. As an example, the intersection with the $y=0.60$ (green) plane is exhibited. Part of the plane below the $y_c$ surface (at the back of the graph) leads to an A victory while the part above (at the front) leads to a B victory.

To illustrate the reversal process driven by Eq. (\ref{y}) I suggest three scenarios which available polls show to be plausible \cite{s1,s2}. Table (\ref{t1}) exhibits these three scenarios from the perspective of critical A voting intentions $I_{ac}$. The first scenario shown has $x= 0.90$ and  $y=0.65$, which yields critical A voting intentions $I_{ac}=0.4194$. Accordingly, an actual A voting intention $I_a=0.42$ leads to an A victory with $v_a=0.5007$ and actual turnout $ T= 0.7550$. The second scenario has $x= 0.90$ and  $y=0.70$, which yields critical A voting intentions $I_{ac}=0.4375$. Accordingly, an actual A voting intention $I_a=0.44$ leads to an A victory with $v_a=0.5025$ and actual turnout $ T= 0.7880$. The final scenario considers $x= 0.85$ and  $y=0.695$, which yields a critical A voting intention $I_{ac}=0.4498$. Accordingly, an actual A voting intention $I_a=0.45$ leads to an A victory with $v_a=0.5002$ and actual turnout $ T= 0.7648$. An additional case still with $x= 0.85$, $y=0.695$ and $I_{ac}=0.4498$ is given to show that  an actual A voting intention $I_a=0.43<I_{ac}=0.4498$ leads to  A loosing with $v_a=0.4799$ and actual turnout $ T= 0.7617$.

\begin{table}
\centering
\begin{tabular}{|c|c||c||c|c|c|} 
  \hline
 A turnout x & B turnout y & Critical $I_{ac}$ & Actual $I_{a}$ & Turnout T& Actual $v_a$\\
  \hline
  0.90 & 0.65 & 0.4194 & 0.42& 0.7550 & 0.5007\\
  \hline
  0.90 & 0.70 & 0.4375 & 0.44 & 0.7880 & 0.5025\\
  \hline
  0.85 & 0.695 & 0.4498 & 0.45 & 0.7648 & 0.5002\\
  \hline
  0.85 & 0.695 & 0.4498 & 0.43 & 0.7617 & 0.4799\\
  \hline
\end{tabular}
\caption{Three cases of A and B turnouts (x and y) are considered. For each one the critical A voting intention $I_{ac}$ is calculated. Then, voting intentions $I_{a}>I_{ac}<\frac{1}{2}$ are shown to yield a voting ballot $v_a>\frac{1}{2}$. Associated turnouts are calculated. Last line shows a case for which the reversal does not occur.}
\label{t1}
\end{table}

These three scenarios can be looked at from the perspective of critical B turnout as shown in Table (\ref{t2}). The first scenario starts with $x= 0.90$ and actual A voting intentions $I_a=0.42$ to yield a critical B turnout $y_c=0.6517$. Actual B turnout $y=0.65<y_c=0.6517$ gives an A victory with $v_a=0.5007$ and actual turnout $ T= 0.7550$. The second reads $x= 0.90$ with actual A voting intentions $I_a=0.44$. This yields a critical B turnout $y_c=0.7071$. Therefore, actual B turnout $y=0.70$ gives an A victory with $v_a=0.5025$ and actual turnout $ T= 0.7880$. The final scenario has $x= 0.85$ and actual A voting intentions $I_a=0.45$ which yields a critical B turnout $y_c=0.6955$. A turnout $y=0.695$ lead to $v_a=0.5002$ and actual turnout $ T= 0.7648$. A fourth scenario still with $x= 0.85$ but with actual A voting intentions $I_a=0.43$ is given. The critical B turnout is $y_c=0.5667$ making $y=0.695>y_c=0.5667$ not allowing the reversal with A loosing at $v_a=0.4799$ and actual turnout $ T= 0.7617$.

\begin{table}
\centering
\begin{tabular}{|c|c||c||c|c|c|} 
  \hline
  A turnout x & Actual $I_{a}$  & Critical $y_c$ & Actual y & Turnout T & Actual $v_a$ \\
   \hline
  0.90 & 0.42 & 0.6517 & 0.65& 0.7550 & 0.5007\\
  \hline
  0.90 & 0.44 & 0.7071 & 0.70 & 0.7880 & 0.5025\\
  \hline
  0.85 & 0.45 & 0.6955 & 0.695 & 0.7648 & 0.5002\\
  \hline
   0.85 & 0.43 & 0.5667 & 0.695 & 0.7617& 0.4799\\
  \hline
\end{tabular}
\caption{Identical three cases as in Table (\ref{t1}) using Eq. (\ref{y}), i.e.,  A turnout x and voting intention $I_a$ are given. The corresponding critical B turnouts $y_{c}$ are calculated. For several actual B turnouts $y<y_{c}$,  T and $v_a$ are calculated. Last line shows a case for which the reversal does not occur.}
\label{t2}
\end{table}

At this stage, before applying  above results to the upcoming 2017 French presidential election, it is worth to notice that during the public campaign which takes place before an election, each candidate tries to gain a maximum number of voting intentions.  It produces a dynamics of public opinion which drives an initial distribution of voting intentions toward a final distribution, which eventually determines the final outcome of the election. Successive polls show how overall support for each candidate evolves during the campaign period. Accordingly, if we consider poll estimates to be exact, in principle a last day poll prior to the election should yield the voting outcome. 

However, such a statement was proven wrong with recent 2016 poll failures to predict in particular the Brexit and Donald Trump election. The failure origin could trace back 
to either a technical drawback related to sample composition and size, or to an unanticipated and sudden shift in individual choices on the very day of the vote. Indeed, it happens that I did predict successfully these two poll breakdowns using my sociophysics model of opinions dynamics  \cite{book}.

For the Brexit I did warn against holding referendums about the European construction many years ago pointing to the likelihood of a rejection despite earlier polls yielding large support to the vote in favor of it \cite{r1,r2}. Along this line I also predicted successfully the 2005 French referendum rejecting the project of European constitution \cite{lehir}. Using the same model I predicted a few months ahead of the vote the totally unexpected victory of Donald Trump at the 2016 US presidential election \cite{trump}. I also forecast the 2002 Le Pen electoral breakthrough \cite{libe}.

All these studies have enlightened the occurrence of non linear phenomena with sudden and abrupt change of individuals choices, supporting the second origin, i.e., sudden opinion shifts, for poll failures. Nevertheless, here I have advocated a third reason, which is not connected to a dynamics of choice shiftings, to turn wrong ``exact" poll estimates. Differentiated abstention, which does modify individual choices, but accounts for the making of a voting intention into casting a ballot, was proven to boost a minority candidate to first place the voting day.

However such a ranking shift requires the existence of a significant gap in respective turnouts for the competing candidates. And it turns out that the 2017 French presidential election will produce such a gap in respective abstentions.

This fact stems from the actual growing reluctance from potential members of the Republican Front to vote for either one of Marine Le Pen leading challengers, Fran\c{c}ois FIllon or Emmanuel Macron. The novelty is to have a simultaneous double reluctance among many individuals, which will create a kind of voting paralysis among committed anti-NF  individuals ending into a solid gap in respective abstentions. 

This differentiation process is expected to be accentuated by the very fact that most Marine Le Pen voters are people who want to vote for her while a good deal of her challenger voters are people who want to oppose her. This asymmetry will contribute in making abstention considerably higher for the challenger than for Le Pen, making above cases with $I_a<\frac{1}{2}$ and $v_a>\frac{1}{2}$ very plausible. For instance, 42\% can lead to 50.07\% as shown in Table (\ref{t1}). Such an outcome is not because of a multi-level system as in the United States but more prosaically because of the discriminated role that abstention will play in the next presidential election.

To conclude, I have used a very simple analysis to show that differentiated abstention can have a drastic effect on an election outcome. In particular, when applied to the second round of upcoming 2017 French Presidential election to be held on May 3, I have proved that for the first time in the National Front history its candidate has a real chance of winning the race to become the next French President despite voting intentions about only 40-45\%.

\end{document}